\documentclass[aps,prl,twocolumn,letterpaper,superscriptaddress,showpacs]{revtex4}
\usepackage{graphicx}
\usepackage{CJK}

\begin{document}

\begin{CJK*}{UTF8}{bsmi}
\title{Unconventional superconducting gap in underdoped cuprates}
\author{Yucel Yildirim}
\affiliation{CMPMSD, Brookhaven National Laboratory, Upton, NY 11973-5000,U.S.A.}
\author{Wei Ku(顧威)}
\altaffiliation{corresponding email: weiku@bnl.gov}
\affiliation{CMPMSD, Brookhaven National Laboratory, Upton, NY 11973-5000,U.S.A.}
\affiliation{Physics Department, State University of New York, Stony Brook, New York 11790, USA}
\date{\today}

\begin{abstract}
A generic theory of the quasi-particle superconducting gap in underdoped cuprates is derived in the strong coupling limit, and found to describe extremely well the experimental ``second gap'' in \textit{absolute scale}.  In drastic contrast to the standard theories of Bogoliubov quasi-particle excitations, the quasi-particle gap is shown to originate from anomalous kinetic process, completely unrelated to the pairing strength.  Furthermore, the $k$-dependence of the gap deviates significantly from the pure $d_{x^2-y^2}$-wave of the order parameter.  Our study reveals a new paradigm for the nature of superconducting gap, and is expected to reconcile numerous apparent contradictions among existing experiments toward a more coherent understanding of high-temperature superconductivity.
\end{abstract}


\maketitle
\end{CJK*}

Recent exciting discovery of the superconducting gap (SCG) in the underdoped cuprates by angle-resolved potoemission spectroscopy (ARPES)~\cite{tg01,Arpes1,tg1,tg2,tg3,tg4,tg5,tg6,tg7,Hufner,Yoshida,Kaminski} and scanning tunneling microscopy (STM)~\cite{STM1,STM2,STM3,GomesSTM4,PasupathySTM5,KohsakaSTM6,BoyerSTM7} reveals critical clues for the puzzling high-temperature superconductivity, which has proven to be one of the most important yet challenging problems of condensed matter physics for more than two decades.  Indeed, unlike the larger pseudo-gap (PG), the SCG closes exactly at the transition temperature $T_c$~\cite{Kaminski}, and shows strong correlation to the doping dependence of $T_c$~\cite{tg8,tg1,tg2,sizeDe}.  Yet, some experiments found intriguing indications that the $k$-dependence of the SCG might not follow the well-established pure $d$-wave of the order parameter~\cite{Kaminski}.  Obviously, a deeper understanding of the properties of the SCG in the quasi-particle (QP) excitation spectra holds the essential key to a resolution to the long-standing problem of high-temperature superconductivity in underdoped cuprates.

Surprisingly, despite the intensive experimental studies, to date there have been limited attempts~\cite{Theory0,Theory1,Theory2,Theory3,Theory4} to understand the SCG, in great contrast to the numerous efforts to address the PG phenomenon~\cite{PG1,PG2}.  Furthermore, the existing understanding for the SCG remains in the scope of the weak coupling BCS framework, which was proposed insufficient~\cite{EmeryKivelson} in the underdoped regime to account for the essential phase fluctuation of the order parameter.  As a result, most of the essential questions remain open, including the energy scale that controls the size of SCG, the true nature of the sharp QP at the edge of the SCG, and the precise $k$-dependence of the SCG, among others.

In this letter, we address these key questions of the SCG by deriving rigorously a generic description in the strong coupling limit, in which real-space pairs of holes are assumed tightly bound in nearest neighboring sites.  With a rigorous separation of Hilbert space into bosonic and fermionic portions that describe the bound pairs and unbound doped holes, respectively, the low-energy fermionic coupling to the Bose condensate is made explicit.  The resulting size of the SCG is shown to describe accurately the experimental gap in absolute scale without any free parameter.  Intriguingly, completely different from the standard Bogoliubov QP excitation, the SCG is found to originate from the anomalous kinetic process unrelated to the pairing strength.  Furthermore, the $k$-dependence of the SCG deviates significantly from the pure $d_{x^2-y^2}$-wave of the underlying order parameter.  The new paradigm fills a long-standing void in our knowledge on the SCG in the opposite limit of the BCS theory, and is expected to reconcile existing and future experiments on the QP SCG toward a comprehensive understanding of high-$T_c$ superconductivity.

Since the SCG occurs at lower energy and lower temperature than the PG physics, we proceed as following.  First, the average kinetics of the fully dressed one-particle propagator are obtained from fitting the experimental measurement of the spectral functions \textit{in the ``normal state'' PG phase} where coherent pairs are not yet formed.  Second, the many-body Hilbert space of the doped holes is split rigorously into those spanned by paired holes (bosons) and the unpaired holes (fermions).  Finally, the QP SCG is obtained from the fermion's coupling to the coherent pairs formed below $T_c$.

\begin{table}[b]
\begin{tabular}{c c c c c}
$\delta(\%)$ & 5.2 & 7 & 15 & 22\\
\hline
$t$ & -4.0 & -3.7 & 1.1 & 8.6\\
$t^\prime$ & 29.8 & 30.6 & 33.5 & 35.1\\
$t^{\prime\prime}$ & 29.8 & 29.0 & 25.8 & 23.9\\
\end{tabular}
\caption{\label{tab:tab1}
Doping dependent hopping parameters (in meV) extracted from ARPES in the PG phase of LSCO \cite{band1,band2}.}
\end{table}

Table I shows the tight-binding parameter obtained from fitting the available ARPES data~\cite{band1,band2} in the PG phase of La$_{2-\delta}$Sr$_{\delta}$CuO$_4$ over a large energy ($\sim$0.7eV) range.  The resulting Hamiltonian, $H=\sum_{ii^\prime} t_{ii^\prime}c_{i}^\dagger c_{i^\prime}$, should be properly understood as a convenient representation of the averaged one-particle kinetic process given by the dressed one-particle propagator under the full renormalization of the many-body interactions.  It certainly does not provide the explicit information of the pairing interaction, nor the decay of the QP.  As clearly shown in Table I, the fully dressed band structure is doping dependent, reflecting the competition between the kinetic and potential energy.
The nearest neighbor hopping $t$ is almost entirely suppressed, understandable from the strong anti-ferromagnetic correlation, and is thus dropped from now on.  Interestingly, the value of the second and the third neighbor hopping, $t^\prime$ and $t^{\prime\prime}$, approaches the same near the 5.2\% quantum critical point (QCP) where superconductivity ceases to exist even at temperature $T=0$.  The equivalence of $t^\prime$ and $t^{\prime\prime}$ turns out to have profound influences on the behavior of the superfluid (c.f: Ref~\cite{1stPaper}) and the SCG as shown below.

Next, consider the strong coupling regime where the binding of pairs are stronger than the fully renormalized kinetic energy, such that at low temperature and low energy, the paired holes cannot break apart easily.  Such regime can result from the above suppression of the kinetic process or the large potential energy associated with the anti-ferromagnetic correlation~\cite{AF} and/or the formation of bi-polarons~\cite{Bipolaron}.  (As shown below, the detail of pairing mechanism is actually not essential for the low-energy physics of SCG in this case.)  Consistent with the pure $d_{x^2-y^2}$ symmetry of the order parameter, in such a strong coupling regime, the paired holes are expected to remain primarily first-neighbor to each other as real-space pairs, since doubly occupying a site by doped holes is unlikely in a lightly doped Mott insulator.  Similarly, the probability of finding unpaired holes below or around $T_c$ is small.  It is thus convenient to split the many-body Hilbert space into two components: one spanned by the tightly bound pairs of nearest neighboring holes, $b_{ij}^\dagger = c_{i \uparrow}^\dagger c_{j \downarrow}^\dagger$, and the other by the unpaired holes, $f_{i\sigma}^\dagger$.  For simplicity, a strong anti-ferromagnetic correlation between the nearest neighboring doped holes is assumed here (well justified in the underdoped regime).

\begin{figure}
\includegraphics[width=7cm]{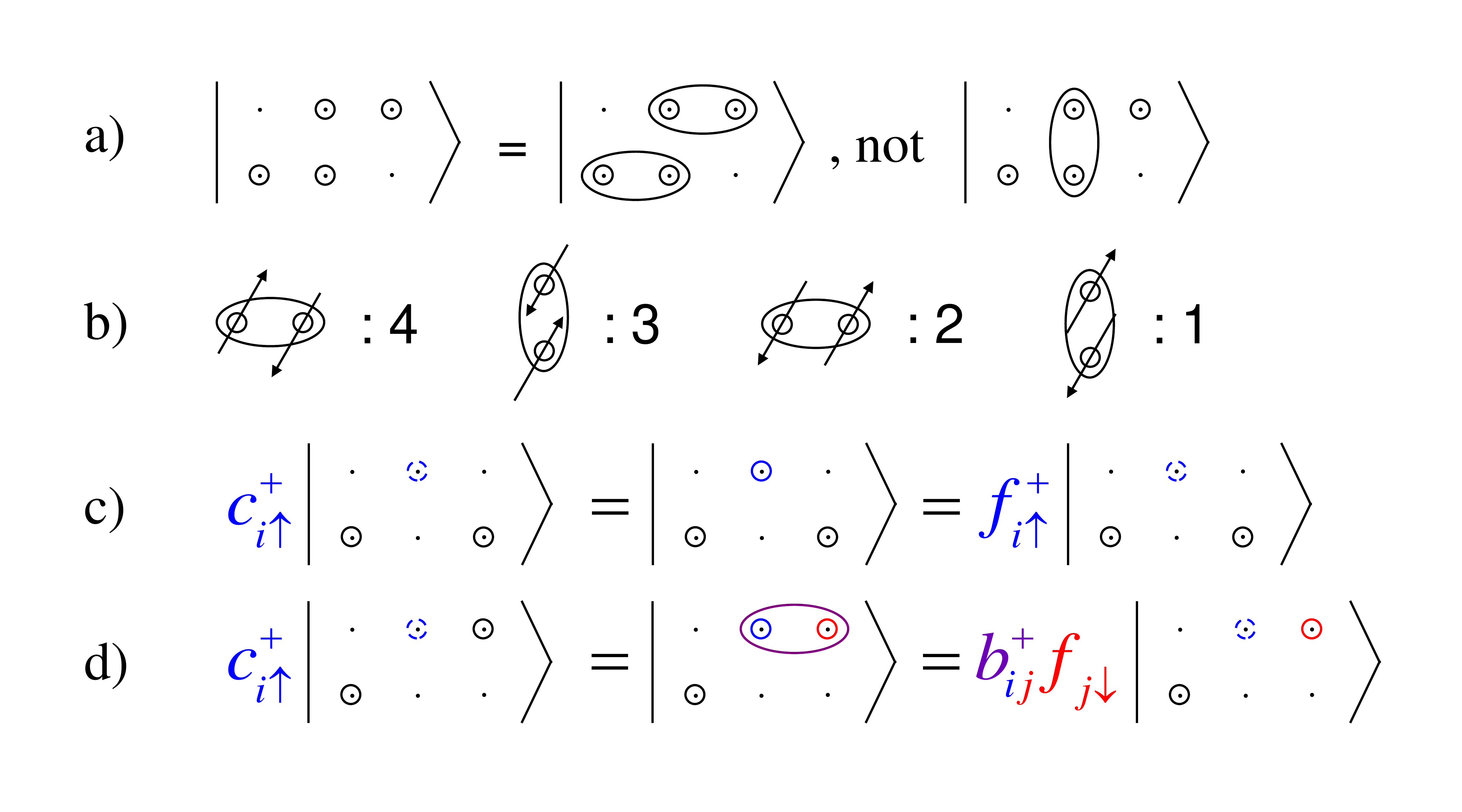}
\caption{\label{fig:fig1} 
(Color online) Illustration of (a) convention (A) and (b) preference factor for a unique representation.  (c)(d) Illustration of two possible contributions of hole creations.}
\end{figure}

Obviously, the two-site nature of the bosons introduces potential ambiguities in the representation.  To ensure a unique representation, the following two conventions are introduced: (A) whenever possible, bosonic representation is employed prior to the fermionic one (c.f.: Fig.~\ref{fig:fig1}(a)), and (B) if multiple representations still exist, the one that maximizes the total ``preference factor'' is employed, where the preference factor is assigned to each pair according to the convention defined in Fig.~\ref{fig:fig1}(b).

In the new representation, the original creation operator of a doped hole now corresponds rigorously to
\begin{eqnarray}\label{eq:eq1}
c_{i\uparrow}^\dagger = f_{i\uparrow}^\dagger + \sum_{j\in NN(i)}^\prime b_{ij}^\dagger f_{j\downarrow} + [ b^\dagger b^\dagger bf + ...] \nonumber\\
c_{j\downarrow}^\dagger = f_{j\downarrow}^\dagger - \sum_{i\in NN(j)}^\prime b_{ij}^\dagger f_{i\uparrow} + [b^\dagger b^\dagger bf + ...]
\end{eqnarray}
where $\sum^\prime$ denotes "sum only the first non-zero" according to the convention defined above.  Figure~\ref{fig:fig1}(c)(d) illustrates the meaning of this non-standard expression.  When creating an additional hole, one can either (first term) create an unpaired hole if and only if no other unpaired hole is next to it (Fig.~\ref{fig:fig1}(c)), or (second term) meet an unpaired hole and form a new pair (Fig.~\ref{fig:fig1}(d)).  It is important to emphasize the mutual exclusion of the terms in Eq.~(\ref{eq:eq1}) according to the above conventions~\cite{noteProjection}.
Owing to the low-density of the doped holes, the higher order terms can be dropped safely.  For a cleaner presentation, the spin indices and the associated sign changes will be suppressed from now on, and addition of hermitian conjugate will be implied.

The effective one-particle kinetics then turns into:
\begin{eqnarray}\label{eq:eq2}
H &=& H^b + H^f + H^{bf}\nonumber\\
&=& (\sum_{ii^\prime j} t_{iji^\prime j}^{b} b_{ij}^\dagger b_{i^\prime j})
 + (\sum_{ii^\prime} t_{i i^\prime}f_{i}^\dagger f_{i^\prime})\\
&+& (\sum_{ii^\prime}\sum_{j}^\prime t_{iji^\prime}^{bf} b_{ij}^\dagger f_{j} f_{i^\prime}
+ \sum_{ii^\prime}\sum_{jj^\prime}^\prime V_{iji^\prime j^\prime}^{bf} b_{ij}^\dagger f_{j} f_{j^\prime} b_{i^\prime j^\prime}),\nonumber
\end{eqnarray}
where $H^{b}$ describes the pivoting motion of the pairs that results in the local $d$-wave symmetry~\cite{1stPaper},
$H^f$ gives the motion of unpaired holes, 
and $H^{bf}$ describes the coupling between the bosons and the fermions.
Obviously, at $T > T_c$ the one-body propagator $G(t,0)\equiv\langle c(t)c^\dagger(0)\rangle = \langle f(t)f^\dagger(0)\rangle + \langle f^\dagger(t)b(t)f^\dagger(0)\rangle + \langle f(t)b^\dagger(0)f(0)\rangle + \langle f^\dagger(t)b(t)b^\dagger(0)f(0)\rangle$ recovers the original experimental dispersion by construction.  (Here $\langle \rangle$ denotes time-ordered ensemble average.)  At low temperature, the last three terms of $G$ are insignificant at low-energy considering 1) the large pairing/depairing energy they need, and 2) the insufficient supply of unpaired holes.
This leaves only the first term in $G$ essential to the low-energy physics of the QP, corresponding to injection of an unpaired hole followed by removal of the same at some other time.

At $T < T_c$, the formation of superfluid of coherent pairs introduces additional effects to the low-energy fermionic Hilbert space through the $t^{bf}$ terms.  Indeed, converting this term to momentum space with a $d_{x^2-y^2}$ phase of the order parameter $\langle b_{k,k^\prime=-k}^\dagger\rangle$, and take the average of $\sum^\prime\rightarrow {1\over 4}\sum$, one obtains $\sqrt{n_0}\sum_k t_k^{bf} f_k f_{-k} + c.c.$ (where $n_0$ denotes the number of pairs per Cu atom in the condensate and can be practially approximated by the superfluid density $n_s$ in number/Cu) with
\begin{eqnarray}\label{eq:eq3}
t_{k}^{bf}&=&{1\over 2}(\cos(k_{x})-\cos(k_{y})) \\
&\times&[4t^\prime \cos(k_{x}) \cos(k_{y}) + 2t^{\prime\prime} (\cos(2k_{x})+\cos(2k_{y}))].\nonumber
\end{eqnarray}
giving the $k$-dependence of the coupling.  The standard Bogoliubov transformation then leads to our main result on the low-energy QP gap:
\begin{eqnarray}\label{eq:eq4}
\Delta_k(T)=t_{k}^{bf}\sqrt{n_0(T)}.
\end{eqnarray}

\begin{figure}
\includegraphics[width=8.5cm]{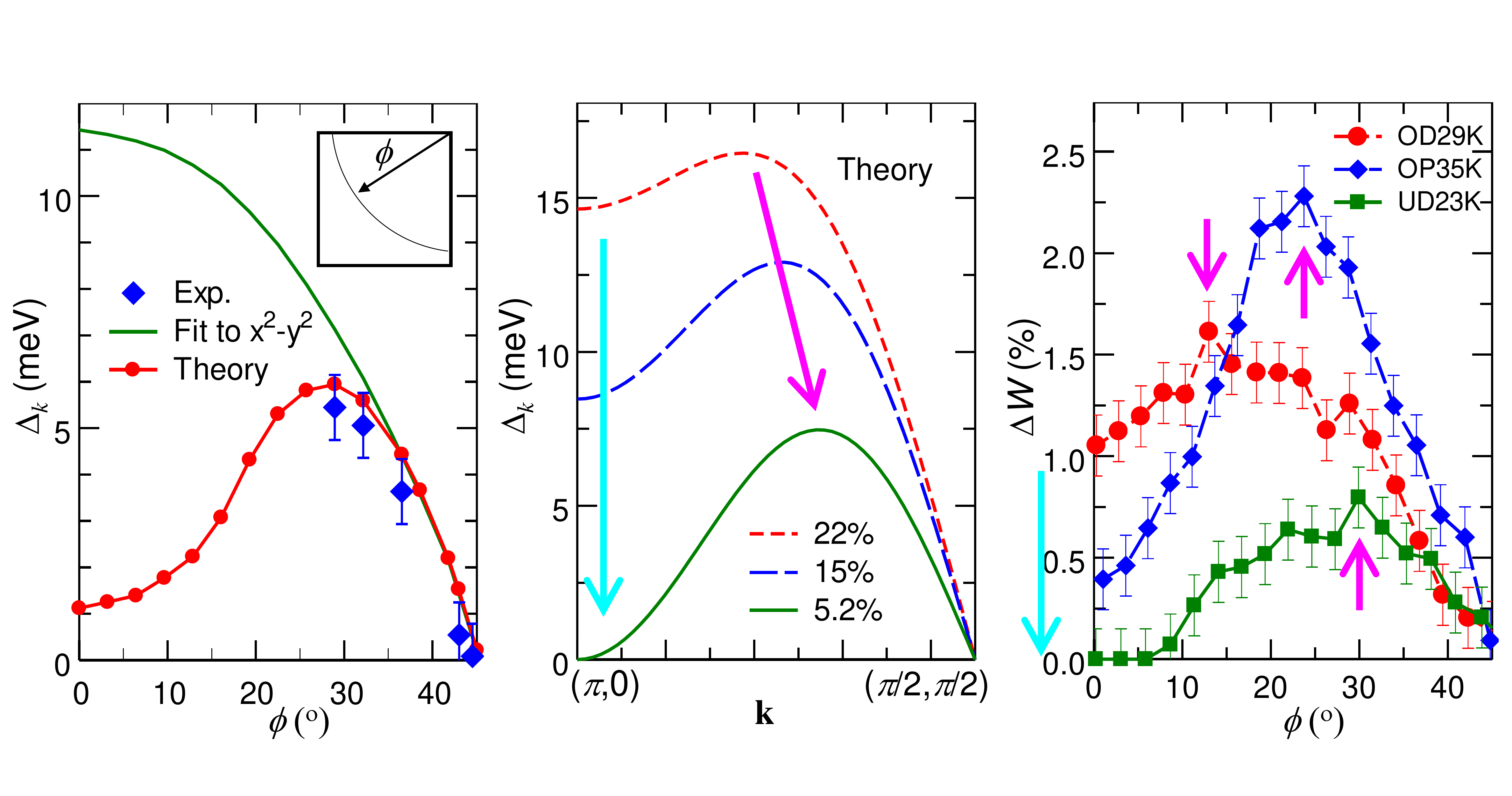}
\caption{\label{fig:fig2}
(Color online) (a) Momentum dependence of the SCG for 7\% doped LSCO at 7.6K.  (b) Theoretical $k$-dependence of the SCG from overdoped (OD/22\%), optimally doped (OP/15\%) to underdoped (UD/5.2\%) systems.  (c) ARPES normalized weight transfer ($\Delta W\propto\Delta_k$) in Bi2201.  Blue and pink arrows indicate the trends upon reduced doping.
}
\end{figure}

Several striking features emerge from this rigorous and generic derivation of the QP SCG in the strong coupling regime.  First, in drastic contrast to the weak coupling BCS theory, the SCG is entirely controlled by the strength of the anomalous kinetic process, $t^{bf}$, and \textit{completely unrelated to the pairing strength!}  To verify this surprising conclusion, we compare in Fig.~\ref{fig:fig2}(a) our result directly with the ARPES~\cite{Yoshida} experimental SCG of La$_{2-\delta}$Sr$_{\delta}$CuO$_4$ (LSCO) at $T=7.6K\sim 0.54T_c$ ($T_c=14K$) for $\delta\sim7\%$.  The theoretical $n_s(T)$ is estimated from $\theta\gamma{\delta\over 2}$, where the thermal reduction factor $\theta(T)=n_s(T\sim 0.54T_c)/n_s(0)\sim0.72$ is directly taken from the penetration depth measurement of YBCO at the same doping level~\cite{Hardy}.  (We are unaware of similar measurement for LSCO.)  The quantum reduction factor $\gamma$ accounts for the drastic suppression of the superfluid density in the vicinity of the 5.2\% QCP, but away from the QCP it should approaches 1 from below rapidly.  Here $\gamma\sim 0.85$ is taken from Ref~\cite{1stPaper} for $\delta=7\%$.

Clearly, the direct comparison in Figure~\ref{fig:fig2} shows an excellent agreement with the experimental SCG \textit{in absolute scale}.  Considering that not even a single free parameter is needed in the theory, this degree of agreement in the size of SCG is truly remarkable.  Evidently, the SCG measured by ARPES and STM (and indirectly by other inelastic measurements) indeed reflects the strength of the effective kinetics of the doped holes, but is completely unrelated to the pairing strength, a feature unimaginable from the weak coupling BCS theory.


The $\delta$- and $k$-dependence of the SCG holds another surprise in our result.  As already hinted near the end of the experimental data set ($\phi\sim 30^\circ$) in Fig.~\ref{fig:fig2}(a), the $k$-dependence of the theoretical SCG deviates dramatically from that of the underlying pure $d_{x^2-y^2}$ symmetry of the order parameter (presented as a fit via the green line).  Unexpectedly, the size of the SCG actually reduces near the anti-nodal direction, instead of reaching its maximum.  A more clear picture is demonstrated in Fig.~\ref{fig:fig2}(b) for three doping levels, $\delta=22\%$ (overdoped), 15\% (optimally doped), and 5.2\% (extremely underdoped).  For a clearer illustration of the trend, $n_0$ is simply set to $\delta / 2$.  (Thus, an exaggeration of the overall scale of the SCG for the overdoped case is to be understood.)

Two apparent trends can be easily identified from Fig.~\ref{fig:fig2}(b).  First, the relative size of the SCG at ($\pi$,0) decreases with reduced doping level.  In fact, Eq.~\ref{eq:eq3} indicates that near the QCP at $\delta\sim 5.2\%$, where $t^\prime$ and $t^{\prime\prime}$ approaches the same value (c.f.: Table I), the size of the SCG reduces to exactly zero at ($\pi$,0), generating additional nodal points.  Second, the $k$-point where the SCG reaches its maximum on the Fermi surface approaches ($\pi/2$,$\pi/2$) as the doping level decreases, in a manner similar to the end of the Fermi arc.

Amazingly, both trends are observed in the recent ARPES analysis~\cite{Kaminski} of the normalized spectral weight transfer $\Delta W$ from $T>T_c$ to $T\ll T_c$ in (Bi,Pb)$_{2}$(Sr,La)$_{2}$CuO$_{6+\delta}$, shown in Fig.~\ref{fig:fig2}(c).  (The properly normalized spectral weight transfer is roughly proportional to the size of the SCG and is likely a much more reliable measure of the SCG than an artificial assignment of a gap edge, given the limited experimental resolution and the very low residual spectral weight in the large PG near ($\pi$,0).)  Indeed, other than the expected exaggeration of the overall magnitude of the over-doped case, our theoretical results resemble very much the experimental ($\delta$,$k$)-dependence, especially capturing the two apparent trends unexpected from a pure $d_{x^2-y^2}$ symmetry of the underlying order parameter, another feature inconceivable from the weak coupling BCS theory.

\begin{figure}
\includegraphics[width=8.7cm]{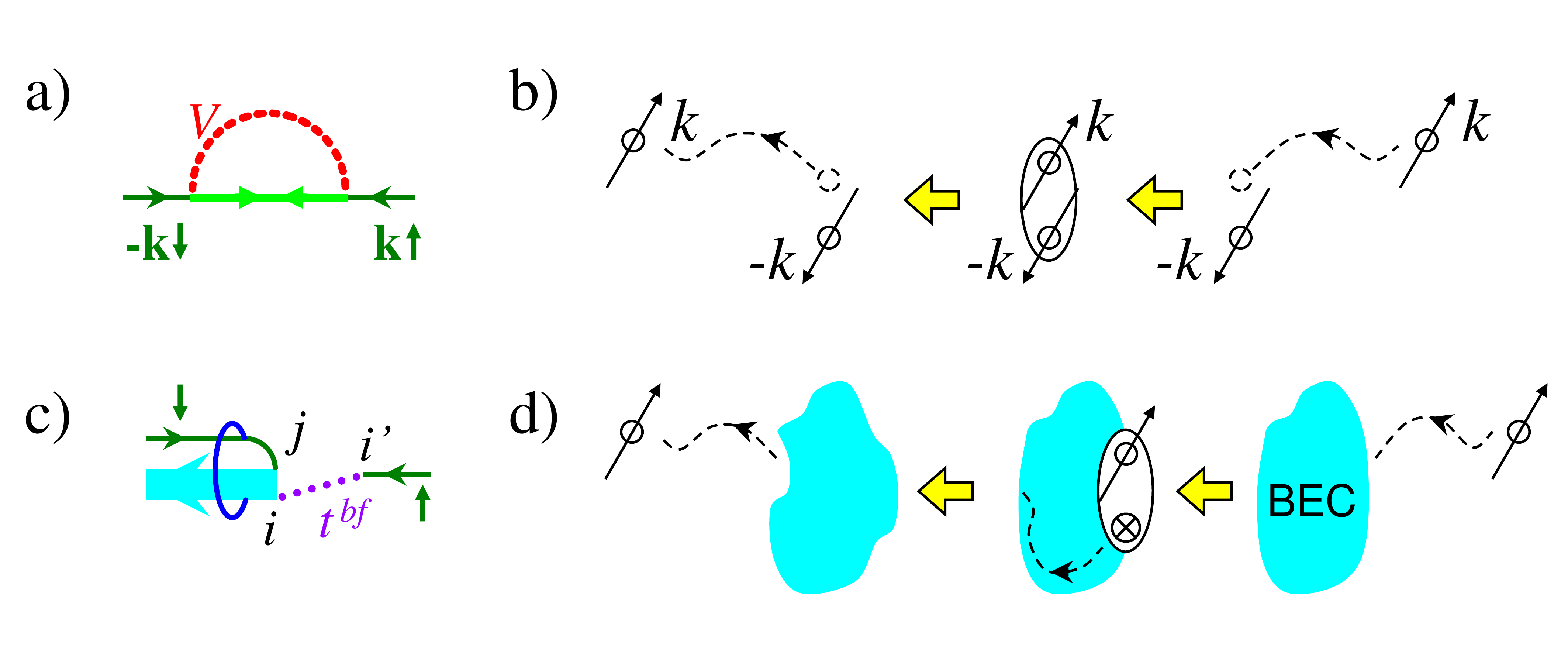}
\caption{\label{fig:fig3}
(Color online) (a) An example of anomalous coupling via pairing interaction $V$ in the BCS theory, with green lines denoting $G$.  (b) Illustration of Bogoliubov QP.  (c) The anomalous kinetic coupling in this work, with thick blue line denoting bosonic propagator, and thin blue ring emphasizing the anti-hole's strong binding to the bosonic space.  (d) Illustration of low-energy QP not involving pairing energy scale.  
}

\label{Int}
\end{figure}

These ``anti-BCS'' features found in our resulting SCG only reflect a new paradigm of the QP SCG in the strong coupling limit, as revealed in our derivation.  Indeed, the smaller SCG observed inside the PG has a novel character entirely different from the well-understood Bogoliubov QP excitation in weak coupling theories (c.f.: Fig.~\ref{fig:fig3}(a)(b)) that breaks apart a particle from the pair with an energy scale of the binding energy.  The coupling $t_{ij}^{bf} b_{ij}^\dagger f_{j} f_{i^\prime} = t_{ij}^{bf} (g_j b_{ij})^\dagger f_{i^\prime}$ allows also another low-energy anomalous kinetic process (Fig.~\ref{fig:fig3}(c)(d)) of an unpaired hole $f^\dagger$ hopping and morphing into an anti-hole $g^\dagger\equiv f$ inside the Bose condensate (BEC), and its hermitian conjugate allows the reverse process of anti-hole hopping into a hole outside the BEC.  Since the creation of the composite object $(g_j b_{ij})^\dagger$ does not involves the energy of the pairing strength (unlike the creation of a single $b^\dagger$), the whole effective scattering process (Fig.~\ref{fig:fig3}(c)(d)) only reflects the energy scale of the kinetics $t^{bf}$.

An important implication of the above new paradigm is that the QPs around the SCG are distinct from the holes that form the pairs.  Thus, they cannot participate the decay of pairs via the standard depairing process~\cite{PatrickLee} advoated previously to explain the near linear reduction of the superfluid density against temperature~\cite{Hardy}.  This leaves thermal phase fluctuation~\cite{EmeryKivelson} or thermal depletion of the BEC~\cite{1stPaper} the only known explanation.  Furthermore, even though both $T_c$~\cite{1stPaper} and SCG are unrelated to the pairing strength, they are correlated via $n_s(0)$: $T_c\propto n_s(0)$~\cite{EmeryKivelson,1stPaper} and $\Delta\propto \sqrt{n_s(0)}$, again defying $\Delta/T_c\sim 1.8$ from the BCS theory.

Obviously, the novel nature of the SCG suggests strongly revisiting the existing theoretical considerations and experimental interpretations of the SCG.  In particular, the decoupling of the $k$-dependence of SCG from that of the order parameter raises an alarming flag on the common practice of equating the nodes in the SCG to those of the order parameter.  Also, the above generic considerations can be applied to the analysis of other strong interaction theories~\cite{tJ,CDMFT} to provide further physical insights unapparent from their numerical solutions.


In conclusion, we develop a generic theory of SCG in the strong coupling limit suitable for underdoped cuprates.  The resulting SCG is shown to differ fundamentally from the standard Bogoliubov QP excitation in the BCS theory.  Not only its size is controlled by the anomalous kinetic process completely unrelated to the pairing strength, but also its ($\delta$,$k$)-dependence deviates significantly from the $d_{x^2-y^2}$-wave of the underlying order parameter.  The new paradigm fills the long-standing void in our knowledge on the SCG in the opposite limit of the BCS theory, and reveals essential nature of the high-$T_c$ superconductivity.

We thank A. Kaminsky for providing the experimental data in Fig.~\ref{fig:fig2}(c).  This work was supported by the U.S. Department of Energy, Office of Basic Energy Science, under Contract No. DE-AC02-98CH10886.


\begin{thebibliography}{10}
\bibitem{noteProjection}
It might be convenient to keep in mind the projection nature of $f_{i}^\dagger\sim Q_{i}^{b} c_{i}^{\dagger} P_{i}^{f}$).  Here $P_i^f\equiv \prod_{j\in NN(i)}(1-n_{j}^f)$ and $Q_{i}^b\equiv \prod_{j\in NN(i)}(1-n_{ij}^{b})$ are projection operators that ensure a unique representation, with $n_{j}^f = f_{j}^\dagger f_{j}$ and $n_{ij}^{b}=b_{ij}^{\dagger} b_{ij}$ being the fermion and boson number density, respectively.
\end{thebibliography}
\end{document}